\documentclass[11pt]{article}

\usepackage{varioref}
\usepackage{amsmath,amsfonts,amsthm,mathrsfs}

\usepackage[normalem]{ulem}
\usepackage{multirow,color,graphics}
\usepackage{amsmath}
\usepackage{amsfonts}
\usepackage{amssymb}
\usepackage{jheppub}
\usepackage{enumerate}
\usepackage{fancyvrb}
\usepackage{verbatim}
\usepackage{wrapfig}
\usepackage{appendix}
\usepackage{amstext}
\usepackage{amssymb}
\usepackage{graphicx}
\usepackage{color}
\usepackage{varioref}
\usepackage{multirow,graphics}
\usepackage{epstopdf}
\newcommand{\nn}{\nonumber}

\numberwithin{equation}{section}

\def\[{\left[}
\def\]{\right]}
\def\({\left(}
\def\){\right)}

    \newcommand{\beq}{\begin{equation}}
    \newcommand{\eeq}{\end{equation}}
    \newcommand\beqa{\begin{eqnarray}}
    \newcommand\eeqa{\end{eqnarray}}
\newcommand\bea{\begin{array}}
\newcommand\eea{\end{array}}

\newcommand{\eq}[1]{(\ref{#1})}

\makeatletter
     \@ifundefined{usebibtex}{} {}
\makeatother

\newcommand{\bra}[1]{\langle #1 |}
\newcommand{\ket}[1]{| #1 \rangle}
\newcommand{\cN}{\mathcal{N}}
\newcommand{\mC}{\mathbb{C}}

\newcommand{\rx}{\mathrm{x}}

\newcommand{\str}{{\rm {str}}\;}

\newcommand{\Ber}{{\rm {Ber}}}

\newcommand{\gt}
{\mathcal T}

\newcommand{\msu}{{\mathfrak{ su}}}

\newcommand{\Bg}{B^{g}}

\newcommand{\cB}{\mathcal{B}}
\newcommand{\cF}{\mathcal{F}}

\title{New  Compact Construction of Eigenstates for Supersymmetric  Spin Chains}

\author[a,b]{~~Nikolay Gromov,}
\author[c,1]{~~Fedor Levkovich-Maslyuk\note{Also at Institute for Information Transmission Problems, Moscow 127994, Russia}}

\affiliation[a]{Mathematics Department, King's College London,
The Strand, London WC2R 2LS, UK}
\affiliation[b]{St.Petersburg INP, Gatchina, 188 300, St.Petersburg,
  Russia}
\affiliation[c]{Departement de Physique, Ecole Normale Superieure / PSL Research University, CNRS, 24 rue Lhomond, 75005 Paris, France }

\emailAdd{nikgromov$\bullet$gmail.com}
\emailAdd{fedor.levkovich$\bullet$gmail.com}

\abstract{
The problem of separation of variables (SoV) in supersymmetric spin chains is closely related to the calculation of correlation functions in ${\cal N}=4$ SYM theory which is integrable in the planar limit. 
To address this question we find a compact formula for the spin chain eigenstates, which does not have any sums over auxiliary roots one usually gets in the widely adopted nested Bethe ansatz. 
Our construction only involves one application of a simple $B^{g}(u_k)$ operator to the reference state for each of the magnons, in complete analogy with the $\msu(2)$ algebraic Bethe ansatz.
This generalizes our SoV based construction for $\msu(n)$ to  the supersymmetric $\msu(1|2)$ case. 
}

\begin{document}

\maketitle

\newpage

\section{Introduction}
\label{sec:intro}

Quantum spin chains serve as a prototypical example of an integrable model combining remarkable algebraic structures with physical relevance. Their supersymmetric versions based on $\msu(m|n)$ superalgebras have also been extensively studied and appear in a wide range of contexts from condensed matter \cite{Sutherland1, Hubbard1} to integrable AdS/CFT \cite{Beisert:2010jr}. While the spectrum of integrable spin chains is typically governed by a concise system of Bethe equations, it is much more difficult to explicitly construct the eigenstates of the spin chain Hamiltonian. The construction of eigenstates is relevant in particular for the calculation of correlators in $\cN=4$  SYM which is actively being explored (see e.g. \cite{Bargheer:2017nne,Komatsu:2017buu,Basso:2015zoa,Fleury:2016ykk,Cavaglia:2018lxi,Giombi:2018qox}).

For the simplest spin chains with a rank-1 symmetry algebra such as $\msu(2)$ or $\msu(1|1)$, one can efficiently build the states via algebraic Bethe ansatz, by repeatedly acting on the vacuum  with a single `creation' operator $B(u)$, 
\beq
\label{psiBB}
    \ket{\Psi}=B(u_1)B(u_2)\dots B(u_K)\ket{0}
\eeq
 where $u_i$ are the Bethe roots which define the excitations' momenta in the spin chain. However, for higher rank spin chains, the standard construction of eigenstates is much more involved. In particular, in the standard nested Bethe ansatz approach the problem is solved recursively, by reducing it to the solution of simpler spin chains with lower  rank symmetry \cite{SutherlandN1,Kulish:1983rd,Belliard:2008di}. The resulting expression for the eigenstate is a complicated sum in which the  number of terms grows exponentially with the number of excitations\footnote{Other remarkable  constructions are known, but they are also rather complicated and typically suffer from exponential complexity as well, see e.g. \cite{Belliard:2012sn,Tarasov1994,Currents,BelPakR10,Albert:2000ne} and the review \cite{Pakuliak:2018rdx}. The eigenstates problem has also been discussed in a pure mathematics context, see e.g. the recent works \cite{Aganagic:2017gsx}. }.
%This approach is equally applicable to the $\msu(n)$ or $\msu(m|n)$ models.

Surprisingly, it was recently  realized in \cite{Gromov:2016itr} that for rational $\msu(n)$ spin chains it is possible to completely bypass this standard recursive  procedure. In fact for any $\msu(n)$  one can build an operator $B^{g}(u)$ which\footnote{In \cite{Gromov:2016itr} this operator was denoted by $B^{\rm good}$. Here we use the shorter notation $B^g$, also utilized in \cite{Liashyk:2018qfc}.} generates the states just as in the simplest $\msu(2)$ case, by repeated action on the vacuum state:
\beq
\label{psiBg}
     \ket{\Psi}=B^{g}(u_1)B^{g}(u_2)\dots B^{g}(u_K)\ket{0} \ .
\eeq
This operator $B^{g}$ is an explicit polynomial in the monodromy matrix entries, and it is the same for any spin chain length and number of excitations. The parameters $u_i$ in \eq{psiBg} are the momentum-carrying Bethe roots fixed by standard nested Bethe equations or by the Baxter equation. Thus, instead of a complicated nested sum the eigenstate is simply given by one term \eq{psiBg}.  For spin chains in the fundamental representation of $\msu(n)$ the construction was extensively checked numerically and proven in several special cases \cite{Gromov:2016itr}. For $\msu(2)$ spin chains it already has nontrivial aspects which were explored further in \cite{Belliard:2018pie} (see also \cite{Belliard:2018pvg} and \cite{FioravNew}, \cite{Sklyaninold}). Very recently, and with remarkable effort, it was proven rigorously for $\msu(3)$ in \cite{Liashyk:2018qfc}, and was also shown there to work for any symmetric representation of $\msu(3)$ on the spin chain sites.

In this paper we show how to extend this highly compact construction of eigenstates to the supersymmetric case. We focus on the first nontrivial example of a higher rank super spin chain, which corresponds to the $\msu(1|2)$ superalgebra \cite{Gohmann:1999av,Essler:1992he,Foerster:1992uk,Gohmann:2001wh}. The $\msu(1|2)$ spin chains are important from a physical point of view as they describe the supersymmetric t--J model widely studied in the context of superconductivity \cite{ZhangCM,SchlottmannCM}. At the same time, they are interesting conceptually due to their intermediate place in complexity between $\msu(2)$ and $\msu(3)$ models. In particular, the challenging problem of finding compact expressions for scalar products of $\msu(3)$ Bethe states \cite{Belliard:2012av,Belliard:2012pr,Belliard:2012is,Pakuliak:2014ela,Pakuliak:2015qga,Pakuliak:2015fma,Slavnov:2015qoa} (e.g. finding an analog of the remarkable Slavnov determinant \cite{SlavnovDet}) seems to be much more tractable in the $\msu(1|2)$ case \cite{Slavnov:2016nhu,Hutsalyuk:2016ndz,Hutsalyuk:2016yii,Hutsalyuk:2016jwh,Hutsalyuk:2016gkn,Hutsalyuk:2017tcx}.\footnote{Related results for general $\msu(m|n)$ spin chains were obtained in  \cite{Pakuliak:2016bhc,Hutsalyuk:2016srn,Fuksa:2017jbl,Hutsalyuk:2017way,Hutsalyuk:2017tcx}. }

We propose an explicit expression for the $B^{g}$ operator which allows one to build the states simply by repeatedly acting on the vacuum, as in \eq{psiBg}. In contrast to the $\msu(n)$ case studied in \cite{Gromov:2016itr}, here $B^{g}$ is not even a polynomial in the monodromy matrix entries. This makes the difference between our construction and the usual nested Bethe ansatz even more striking. While in the bosonic $\msu(n)$ case $\Bg$ is written in terms of certain determinants (quantum minors) built from the monodromy matrix, here we find they should be replaced by  Berezinians which are known to be non-polynomial. At the same time, by a simple redefinition of the monodromy matrix (mutiplication by an explicit scalar function) we can still render the $B^g$ operator a polynomial in $u$ for the spin chain we consider, and its degree\footnote{With a more general twist the degree could likely become $2L$, which is the same as for the $\msu(3)$ operator $B^g$ once we extract a trivial factor from it \cite{Gromov:2016itr}. See the discussion in section \ref{sec:su12in}.} is~$2L-1$.

The construction of \cite{Gromov:2016itr} is directly related to Sklyanin's separation of variables (SoV) program \cite{Sklyanin:1992sm, Sklyanin:1995bm}, which consists of finding special variables in which the dynamics of a many-particle integrable system decouples into a set of non-interacting one dimensional models. In fact eigenstates of the same operator $B^{g}$ provide the basis of separated coordinates for the  $\msu(n)$ spin chain (see also \cite{Sklyanin:1992sm}). Factorization of the wavefunctions in this basis follows immediately from the construction \eq{psiBg} of eigenstates \cite{Gromov:2016itr}. However, in the supersymmetric case it is not known how to obtain the separated variables even for the simplest $\msu(1|1)$ models. Thus it is all the more nontrivial that a direct analog of  the formula for eigenstates  \eq{psiBg}  exists for higher rank super spin chains. 

Implementation of the SoV for supersymmetric models remains an important future goal, especially in view of its relevance for $\cN=4$ SYM where the symmetry algebra is ${\mathfrak{psu}}(2,2|4)$. In the simpler $\msu(1|1)$ case we managed to overcome some of the obstacles towards SoV and we present these results in appendix \ref{sec:su11}. Namely, we propose a $B^{g}$ operator for  $\msu(1|1)$  which, although it does not give separated variables,  is diagonalizable unlike the standard $B$, while having several other curious properties.

This paper is organized as follows. In section \ref{sec:rev} we present our notation and overview of supersymmetric spin chains, and also discuss briefly the $\msu(1|1)$ case. Section \ref{sec:su12in} contains our main results, namely the construction of eigenstates for $\msu(1|2)$ spin chains using only a single operator $\Bg$. We also present the construction for the spin chain with the $(2|1)$  choice of grading, which, although similar to the $(1|2)$ case, is technically different. We conclude in section \ref{sec:concl}. In appendix \ref{sec:su11} we discuss some observations on the SoV  for  $\msu(1|1)$ spin chains, in particular presenting the improved $B$ operator. 

\bigskip
\bigskip
\noindent
\textbf{Note added} 
\\
When finishing the draft we learned about a work in progress \cite{Ltoapp} where related results were obtained independently.

\section{Supersymmetric spin chains overview}

\label{sec:rev}

In this section we review the standard algebraic Bethe ansatz description of super spin chains (see e.g. \cite{Pakuliak:2018rdx,Belliard:2008di} for a review). In the process we introduce  notation used in the rest of the paper.

We will work with graded vector spaces, however we will use only standard complex numbers rather than Grassmann variables. A graded vector space $\mC^{m|n}$ consists of vectors $v$ with components  $v_i\in\mC$ with $i=1,\dots, m+n$. We assign a parity $[i]$ to the indices so that
\beq
    [i]=0 \ \  \text{for}\ \  i=1,\dots,m \ , \ \ \ \ \ \ \ \ [i]=1\ \  \text{for}\ \  i=m+1,\dots,n \ \ \ .
\eeq
This space $\mC^{m|n}$ realizes the fundamental representation of ${\mathfrak su}(m|n)$.
The Hilbert space ${\cal H}$ of the spin chain is a tensor product of $L$ copies of the space $\mC^{m|n}$,
\beq
\label{hilb}
	{\cal H}=\mC^{m|n}\otimes\mC^{m|n}\otimes\dots \otimes\mC^{m|n}\ .
\eeq

The algebraic construction of an integrable quantum spin chain is based  on an R-matrix. The standard rational R-matrix in the supersymmetric case acts on $\mC^{m|n}\otimes\mC^{m|n}$ and is given by\footnote{The R-matrix is  ${\mathfrak{gl}}(m|n)$ invariant, but we will speak about  $\msu(m|n)$ as the symmetry algebra to emphasize that we consider a finite-dimensional representation at each site of the chain.}
\beq
\label{Rgr}
	R^{ij}_{kl}(u)=\delta^i_k\delta^j_l
	+\frac{i}{u}\delta^i_l\delta^j_k(-1)^{[i][j]} \ ,
\eeq
where the extra signs correspond to using a graded permutation operator. Let us also mention that we will often use the notation
\beq
\label{snot}
    f^\pm\equiv f(u\pm i/2), \ \ \ f^{[+a]}\equiv f(u+ia/2) \ 
\eeq
for shifts of the spectral parameter.

Multiplying several R-matrices together we obtain the monodromy matrix $T(u)$ defining the spin chain, which acts in the tensor product of the Hilbert space ${\cal H}$ and an auxiliary space $\mC^{m|n}$,
\beq
\label{TRR}
   T(u)=R_{01}(u-\theta_1)\otimes R_{02}(u-\theta_2)\otimes\dots \otimes R_{0L}(u-\theta_L)\otimes g\ \ .
\eeq
We have introduced an extra twist matrix $g$ which acts in the auxiliary space only and corresponds to twisted boundary conditions. We take it to be diagonal,
\beq
\label{gtw}
	g={\rm diag}\(\lambda_1,\lambda_2,\dots,\lambda_{m+n}\)\ ,
\eeq
and assume that the twists $\lambda_i$ are all distinct and in generic position. They serve as regulators in the construction, and also ensure there is a 1-to-1 correspondence between spin chain states and solutions of the nested Bethe equations. We also introduced the parameters $\theta_L$ which correspond to inhomogeneities of the spin chain\footnote{Although most of the checks we present later in this paper have been done assuming that all $\theta_k$ are in generic position, we expect our results to be valid for any choice of $\theta$'s.}.

The definition of the tensor product of operators  in \eq{TRR} is nontrivial in the supersymmetric case and involves extra signs reflecting the graded nature of the vector spaces $\mC^{m|n}$, see e.g. \cite{Gohmann:1999av}.  Explicitly, the matrix elements of $T(u)$ read
\beqa
	T^{a_1a_2\dots a_L\;j}
	_{a'_1a'_2\dots a'_{L}\;j'}(u)
	&=&
	R^{a_1j}_{a'_1j''}(u-\theta_1)
	R^{a_2j}_{a'_2j''}(u-\theta_2)\dots
	R^{a_Lj}_{a'_Lj''}(u-\theta_L)\;g^{j''}_{j'}\\ \nn&\times&
	(-1)^{\sum_{\alpha=2}^L\sum_{\beta=1}^{\alpha-1}[a'_{\beta}]([a_\alpha]+[a'_{\alpha}])}
\eeqa
where the indices $a_1,a_2,\dots$ and $a'_1,a'_2,\dots$ correspond to individual $\mC^{m|n}$ factors of  the Hilbert space \eq{hilb} while $j,j'$ label the auxiliary space, and we assume summation over repeated indices. 

Although one could consider spin chains with arbitrary representations of $\msu(m|n)$ in the physical and the auxiliary spaces, we only discuss the case when both representations are fundamental. Let us note, however, that other representations on the sites of the chain can be obtained by fusion from the fundamental one, corresponding to a special choice of $\theta$'s.

%\beq
	%T^{\alpha\beta\dots\gamma\rho}_{\alpha'\beta'\dots\gamma'\rho'}(u)=
	%R^{\alpha\rho}_{\alpha'\rho'}(u-\theta_1)R^{\beta\rho}_{\beta'\rho'}(u-\theta_2)\dots
	%R^{\gamma\rho}_{\gamma'\rho'}(u-\theta_L)(-1)^{\sum_{j=2}^L\sum_{i=1}^{j-1}}
%\eeq
%In index-free notation we can write it as
%\beq
	%R_{12}(u)=I+\frac{1}{u}P_{12}^S
%\eeq
%where $P_{12}^S$ is the graded permutation operator.

It is very useful to view $T(u)$ as a matrix of size $(m+n)\times(m+n)$ whose elements $T^i_{j}$ act on the physical Hilbert space ${\cal H}$. This matrix satisfies a graded version of the celebrated RTT relation which in components reads\footnote{The notation we use is slightly different compared to \cite{Hutsalyuk:2016srn,Hutsalyuk:2017tcx}, so that we have
$
	T_{ij}^{\rm there}=(-1)^{[j]([i]+1)}T^{i}_j
$.
Our notation ensures in particular that the Berezinian has the standard form given below in \eq{Ber}, while  the notation of \cite{Hutsalyuk:2016srn,Hutsalyuk:2017tcx} would lead to an extra sign in that expression.}
\beq
\label{RTT}
	{R}^{ji}_{i'j'}(u-v)
	T^{i'}_{i''}(u)
	T^{j'}_{j''}(v) \;(-1)^{[j''][j']}
	=
	T^{i}_{i'}(v)
	T^{j}_{j'}(u) 
	{R}^{j'i'}_{i''j''}(u-v)\;(-1)^{[i'][j]} \ .
\eeq
Below we will not distinguish between $T^i_j$ and $T_{ij}$, in order to write some expressions more concisely.

Another key object is the transfer matrix, defined as the supertrace of the monodromy matrix, 
\beq\label{defstr}
    \gt(u)\equiv\str T(u)=\sum_{i=1}^{m+n}(-1)^{[i]}\;T^i_i(u) \ .
\eeq
One can show that as a consequence of the RTT relation, the transfer matrices form a commutative family,
\beq
   [\gt(u),\   \gt(v)]=0 \ ,
\eeq
which in particular includes the Hamiltonian of the spin chain. Expanding $\gt(u)$ as a series in $u$ one therefore obtains a large set of conserved charges commuting wth the Hamiltonian. The main problem we study in this paper is constructing the common eigenbasis of these operators. A particularly simple eigenvector is given by
\beq
\label{def0}
	\ket{0}=\begin{pmatrix}1\\0\\ \vdots \\ 0  \end{pmatrix}
	\otimes
	\begin{pmatrix}1\\0\\ \vdots \\ 0  \end{pmatrix}
	\otimes\dots \otimes
\begin{pmatrix}1\\0\\ \vdots \\ 0  \end{pmatrix}
\ ,
\eeq
% \beq
% \label{def0}
% 	\ket{0}=\begin{pmatrix}1\\0\\ \vdots \\ 0 \\ \hline 0\\ \vdots \\ 0 \end{pmatrix}
% 	\otimes
% 	\begin{pmatrix}1\\0\\ \vdots \\ 0 \\ \hline 0\\ \vdots \\ 0 \end{pmatrix}
% 	\otimes\dots \otimes
% \begin{pmatrix}1\\0\\ \vdots \\ 0 \\ \hline 0\\ \vdots \\ 0 \end{pmatrix}
% \ ,
% \eeq
and plays the role of the `vacuum'  reference state in the algebraic Bethe ansatz.

% Here the horizontal lines emphasize the $(m|n)$ grading of the vector spaces at each site of the chain.

As an example, let us briefly discuss the $\msu(1|1)$ case which is explored in more detail in appendix \ref{sec:su11}. For $\msu(1|1)$ we may write $T$ as a $2\times 2$ matrix whose entries act on the Hilbert space,
\beq
T(u)=\begin{pmatrix}
	A(u)&B(u) \\ C(u)&D(u)
\end{pmatrix} \ .
\eeq
Then the transfer matrix is given by
\beq
    \gt(u)=A(u)-D(u) \ .
\eeq
The $B$ operator serves as a creation operator generating the eigenstates $\ket{\Psi}$ of the transfer matrix,
\beq
    \ket{\Psi}=B(u_1)B(u_2)\dots B(u_K)\ket{0} \ ,
\eeq
provided $u_i$ are fixed by the Bethe equations which read
\beq
\frac{\lambda_1}{\lambda_2}\prod_{k=1}^L
\frac{u_j-\theta_k+i/2}{u_j-\theta_k-i/2}=1, \ \ j=1,\dots,K
\ \ \ .
\eeq

In the next section we extend this highly compact construction of eigenstates to the higher rank $\msu(1|2)$ spin chains.

\section{Eigenstates for  $\msu(1|2)$ spin chains}

\label{sec:su12in}

In this section we present our main result -- the new  compact construction of eigenstates for  $\msu(1|2)$ spin chains. We first discuss the spin chain with the $(1|2)$ grading in detail and then present the generalization to the $(2|1)$ grading which is similar but technically different. For a pedagogical discussion of the $\msu(1|2)$ algebra and associated spin chains see e.g.  \cite{Frappat:1996pb,Volin:2010cq}\footnote{Mathematical aspects of representations of the corresponding Yangians were discussed in \cite{Zhang:1994ad,Zhang:1995uh}.}.

The construction of states we propose is inspired by \cite{Gromov:2016itr} where it was shown that one can build an operator $B^{g}$ which generates states for $\msu(n)$ spin chains simply by repeated action on the vacuum as in \eq{psiBg}, like in the $\msu(2)$ case. This operator is constructed from quantum minors of the monodromy matrix,  which are defined as determinants of submatrices of $T(u)$ with extra shifts of the spectral parameter $u$. These quantum minors are generalizations of the quantum determinant \cite{MolevBook}. Explicitly, an $n\times n$ quantum minor is given by\footnote{We note that the vertical slash appearing in the l.h.s. of \eq{qmindef} is unrelated to the graded vector space notation such as $\msu(m|n)$.}
\beq\label{qmindef}
	T_{j_1,\dots,j_n|\; k_1,\dots,k_n}(u)=\sum_{\sigma\in S_n}(-1)^{{\rm sign}(\sigma)}
	T_{j_{\sigma(1)}k_1}(u)T_{j_{\sigma(2)}k_2}(u+i)\dots
	T_{j_{\sigma(n)}k_n}(u+ni)\ .
\eeq
With this notation the $\Bg$ operator for the $\msu(3)$ case takes the simple form
\beq
\label{Bsu3}
    \Bg(u)=
     T_{1|3}(u)T_{12|13}(u-i)+T_{2|3}(u)T_{12|23}(u-i)\ ,
     \ \ \text{with}\ \ {T_{ij}\to T^{g}_{ij}}\ ,
\eeq
where we indicated that $T_{ij}$ are substituted by the elements of the improved monodromy matrix which  is defined by
\beq
\label{Tg3}
    T^{g}(u)=K^{-1}T(u)K
\eeq
with $K$ a generic  $3\times 3$ constant
matrix. The extra similarity transformation given by $K$  renders the construction non-degenerate while preserving all commutation relations between entries of $T(u)$. It also leaves unchanged the trace of $T(u)$ and consequently the spin chain Hamiltonian.

Our main observation is that in the $\msu(1|2)$ case one should replace the quantum minors appearing in \eq{Bsu3} by Berezinians  which play the role of determinants for supermatrices \cite{Berezin87,FioresiBook}. For a $2\times 2$ matrix split into four blocks $A,B,C,D$ we define the Berezinian as
\beq
\label{Ber}
	\Ber \begin{pmatrix}
	A(u)&B(u) \\ C(u)&D(u)
	\end{pmatrix}=\(A(u)-B(u)D^{-1}(u)C(u)\)D^{-1}(u) \ .
\eeq
Applying this formula to the monodromy matrix of an $\msu(1|1)$ spin chain  (see \eq{T11}) gives the operator $\Ber\; T(u)$ which is a central element of the Yangian $Y({\mathfrak gl}(1|1))$, i.e. it commutes with $A(v),B(v),C(v)$ and $D(v)$ for all values of $u$ and $v$. This shows that the Berezinian plays the same role for $\msu(1|1)$ as the quantum determinant does for the $\msu(2)$  case. We recall that the quantum determinant for an $\msu(2)$ spin chain monodromy matrix reads
\beq
\label{qdet}
    {\rm qdet} \begin{pmatrix}
	A(u)&B(u) \\ C(u)&D(u)
	\end{pmatrix}=A(u)D(u+i)-C(u)B(u+i)
\eeq
and coincides with the $2\times 2$ quantum minor defined in \eq{qmindef}.

Notice that there are no shifts of the spectral parameter in the $2\times 2$ Berezinian \eq{Ber}, in contrast\footnote{Shifts of $u$ do appear in quantum Berezinians of higher size T-matrices \cite{NazarovBer,Stukopin94,Gow05,Gow07,GowThesis} which generate the center of the Yangian for higher rank spin chains.} to the quantum determinant \eq{qdet}. Due to this we will not make a distinction between the quantum Berezinian and the usual Berezinian of a $2\times 2$ block matrix (both are given by \eq{Ber}).

We will denote the Berezinians similarly to the quantum minors in \eq{qmindef}, namely
\beq
    \Ber_{i_1i_2|j_1j_2}(u)\equiv \Ber
    \begin{pmatrix}
    T_{i_1j_1}(u)&T_{i_1j_2}(u)\\
    T_{i_2j_1}(u)&T_{i_2j_2}(u)
    \end{pmatrix}
     \ .
\eeq
With this notation the $B^{g}$ operator for $\msu(1|2)$ is obtained by simply replacing the quantum minors in the $\msu(3)$ result \eq{Bsu3} by the Berezinians,
\beq
\label{Bg12}
	B^{g}(u)=T_{1|3}(u)\Ber_{12|13}(u)+T_{2|3}(u)\Ber_{12|23}(u)\ \ ,
     \ \ \text{with}\ \ {T_{ij}\to T^{g}_{ij}}\ ,
\eeq
where again one should use elements of the improved monodromy matrix $T^{g}$. It is defined by \eq{Tg3} like in the $\msu(3)$ case,
with the sole difference being that $K$ should only have nonzero entries in its `even' diagonal blocks,
\beq
\label{K12}
    K=\(
    \begin{array}{c|cc}
    K_{11}& 0&0\\
    \hline
    0& K_{22}&K_{23}\\
    0& K_{32}&K_{33}
    \end{array}
    \)\ ,
\eeq
where the lines emphasize the splitting of $K$ into odd and even elements as an operator on the graded vector space $\mC^{1|2}$.
The construction works as long as $K$ is a generic matrix of this type. Let us also note that in contrast to the  $\msu(3)$ case, there are no shifts of $u$ at all in \eq{Bg12}.

The main property of this operator is that it allows one to build the transfer matrix eigenstates just as for $\msu(1|1)$ or $\msu(2)$, by repeated action on the vacuum! Namely,
\beq
\label{psi12}
    \ket{\Psi}=B^{g}(u_1)B^{g}(u_2)\dots B^{g}(u_K)\ket{0} \ .
\eeq
The state $\ket{0}$ here is the standard reference state \eq{def0},
\beq
    \ket{0}=\begin{pmatrix}1\\\hline 0\\  0\end{pmatrix}\otimes\begin{pmatrix}1\\ \hline 0\\  0\end{pmatrix}
	\otimes\dots \otimes \begin{pmatrix}1\\\hline 0\\  0\end{pmatrix}
\eeq
where the horizontal lines again highlight the $(1|2)$ grading of the $\mC^{1|2}$ space at each site. Like for $\msu(3)$, the $u_i$ in \eq{psi12} are the momentum-carrying Bethe roots. One way to fix them is to solve the set of usual nested Bethe ansatz equations, which also include auxiliary roots~$v_j$,
\beqa
\label{bae12}
   \prod_{k=1}^L\frac{u_i-\theta_k+i/2}{u_i-\theta_k-i/2}&=&\frac{\lambda_2}{\lambda_1}\prod_{j=1}^N\frac{u_i-v_j+i/2}{u_i-v_j-i/2}\ , \ \ \ i=1,\dots,K
   \\ \label{bae122}
    \prod_{j=1}^K\frac{v_i-u_j+i/2}{v_i-u_j-i/2}&=&-\frac{\lambda_2}{\lambda_3}\prod_{j=1}^N\frac{v_i-v_j+i}{v_i-v_j-i}\ ,\ \ \ i=1,\dots,N \ .
\eeqa
The only role of the auxiliary roots is that they indirectly affect the values of the main roots $u_i$ through the Bethe equations\footnote{We consider  spin chains with generic twists $\lambda_i$, which lift degeneracies and ensure that the states are in 1-to-1 correspondence with solutions of the Bethe equations.}. Let us also note that in terms of the Bethe roots the eigenvalues of the transfer matrix read (as one can deduce via standard methods)
\beq
\label{T12e}
\gt(u)=Q_\theta^-\[\lambda_1\frac{Q_\theta^{+}Q_u^{--}}{Q_\theta^- Q_u}-\lambda_2\frac{Q_u^{--}Q_v^{+}}{Q_uQ_v^-}
	-{\lambda_3}\frac{Q_v^{---}}{Q_v^-}\] \ ,
\eeq
where
\beq
Q_\theta(u)=\prod_{k=1}^L(u-\theta_k)\ , \ \ \ 
    Q_u=\prod_{j=1}^K(u-u_j) \ , \ \ \ Q_v=\prod_{j=1}^N(u-v_j) \ 
\eeq
and we also used the compact notation \eq{snot}.

% Alternatively one may completely avoid any reference to the nested Bethe ansatz by reformulating the Bethe equations as a 3rd order Baxter equation for the Q-function $Q(u)=\prod_i(u-u_i)$ encoding the momentum-carrying roots \com{to check}. In this approach the auxiliary roots do not enter the equations at all.

Our construction of the states is clearly free from the recursion inherent in the standard nested Bethe ansatz, where the eigenstates are built in terms of the wavefunctions of an auxiliary lower rank spin chain. Our approach involves only a single operator $B^{g}$ acting repeatedly on the vacuum.   Curiously, while in the nested Bethe ansatz states are built by polynomial combinations of monodromy matrix elements acting on the vacuum (with complicated state-dependent coefficients), the $B^{g}$ operator is not even a polynomial in the $T_{ij}$ operators\footnote{Let us note that for $\msu(3)$ one can also write $B^g$ as a non-polynomial combination of $T_{ij}$, using that the $2\times 2$ quantum minors entering \eq{Bsu3} and given by \eq{qdet} can be equivalently written as
$A(u)(D(u+i)-C(u+i)A^{-1}(u+i)B(u+i))$. For $\msu(1|2)$, however, our $B^g$ operator cannot be recast as a polynomial of  $T_{ij}$, as it is not polynomial in $u$.}.
It is also not a polynomial of $u$ in the representation we consider, but we observed that one can make the dependence on $u$ polynomial by multiplying the $T_{ij}$ matrix by a scalar function, namely by replacing $T_{jk}(u)\to Q_\theta(u-\tfrac{i}{2}) T_{jk}(u)$. After that $B^g(u)$ becomes a polynomial of degree $2L-1$. With a more general twist $K$ (e.g. one involving off-diagonal Grassmann entries) it might be possible to make the degree $2L$, i.e. the same as it is for the $B^g$ operator in $\msu(3)$  once we remove from it a trivial overall factor \cite{Gromov:2016itr}.

Let us highlight a peculiar structural feature of the $B^{g}$ operator for $\msu(1|2)$. Although all the monodromy matrix entries $T_{ij}(u)$ are just operators acting on the Hilbert space, it is useful to label them as either bosonic/even ($\cB$) or fermionic/odd ($\cF$) depending on their position inside $T$ viewed as a $(1|2)$ supermatrix, so that schematically
\beq
\label{Tgr}
    T(u)=\left(
    \begin{array}{c|cc} \cB & \cF & \cF
    \\ \hline \cF & \cB & \cB\\
    \cF& \cB&\cB
    \end{array}
    \right) \ .
\eeq
The Berezinian is naturally defined for $2\times 2$ matrices with the standard grading \scriptsize${\left(
    \begin{array}{cc} \cB & \cF 
    \\ \cF & \cB
    \end{array}
    \right)}$\normalsize. While the first Berezinian in the expression \eq{Bg12} for $\Bg$ is indeed applied to a matrix with  this grading, the second one is evaluated for a matrix of the type \scriptsize $\left(
    \begin{array}{cc} \cF & \cF 
    \\  \cB & \cB 
    \end{array}
    \right)$\normalsize. When computing  $B^{g}$ we simply evaluate this Berezinian formally using the definition \eq{Ber}. It would be highly interesting to  understand the algebraic meaning of such non-conventional super determinants.

While the proposed construction of eigenstates \eq{psi12} should be regarded as a conjecture, we have extensively checked it numerically.  We verified  that it produces all the states for the spin chain with $L=1,2,3$ or $4$ sites. We also tested it for several states with up to four excitations for $L=5$, where we already have large $243\times 243$ matrices. In addition, we have proven it analytically for the case with one excitation for any spin chain length $L$, with a particular simple choice of $K$
\beq
	K=\(\begin{array}{c|cc}
		1&0&0\\
		\hline
		0&1&1\\
		0&0&1
	\end{array}\) \ .
\eeq
The proof is essentially by brute force and follows the one for $\msu(3)$ in \cite{Gromov:2016itr}. It is based on the fact that the vacuum is an eigenstate for all elements of $T_{ij}^{g}$ except $T^{g}_{12}$ and $T^{g}_{13}$ which serve as creation operators. Namely, we have \beq
\label{Tan}
	T^{g}_{ij}(u)\ket{0}=0, \ i>j
\eeq
and also
\beq
\label{Td}
	T^{g}_{11}(u)\ket{0}=\lambda_1Q_\theta^+\ket{0}, \ \ 
	T^{g}_{22}(u)\ket{0}=\lambda_2Q_\theta^-\ket{0}, \ \ 
	T^{g}_{33}(u)\ket{0}=\lambda_3Q_\theta^-\ket{0}, \ \ 
\eeq
\beq
\label{T23}
	T^{g}_{23}(u)\ket{0}=(\lambda_2-\lambda_3)Q_\theta^-\ket{0}, \ \ 
\eeq
as one can verify similarly to the $\msu(3)$ case \cite{Gromov:2016itr}\footnote{For the original monodromy matrix elements $T_{ij}(u)$ we have the same action on the vacuum \eq{Tan}, \eq{Td} but instead of \eq{T23} we find that $T_{23}$ annihilates the vacuum.}. 
Then using the RTT relations to commute all $T_{ij}^{g}$ to the right of $T^{g}_{12}$ and $T^{g}_{13}$ until they hit the vacuum state $\ket{0}$, we get
\beq
    \Bg(u)\ket{0}=
    \lambda_2Q_\theta^-
    \(T^{g}_{12}(u)+\frac{\lambda_2Q_\theta^--\lambda_1Q_\theta^+}{(\lambda_2-\lambda_3)Q_\theta^-}T^{g}_{13}(u)\)\ket{0} \ .
\eeq
Acting on this expression with the transfer matrix one can similarly show that it is an eigenstate on the solutions of Bethe equations\footnote{Extending this proof to 2 magnons is already nontrivial due to the need to commute $\(T_{23}^{g}\)^{-1}$ appearing in the Berezinians through other elements of $T^{g}$.}. We leave for the future a full general proof for any number of excitations, and hope it can be done using the recent techniques of \cite{Liashyk:2018qfc}.

% % For the $\msu(m|n)$ spin chains the monodromy matrix naturally has a supermatrix structure, so that elements in its diagonal $m\times m$ and $n\times n$ blocks may be regarded as 'even' and the other elements as 'odd'. 

% Above in section \ref{sec:dual} we mentioned that for $\msu(1|1)$ the $\Bg$ operator we have constructed is not suitable for building the states starting from a different pseudovacuum reference state. The same is true for $\msu(1|2)$ as we find that our operator $\Bg$ annihilates both states that serve as alternative pseudovacua, namely

Let us note that the $B^{g}(u)$ operators for $\msu(1|2)$ do not commute at different values of $u$, and thus naively are not suitable for definition of separated variables (in contrast to the $\msu(n)$ case \cite{Gromov:2016itr}). Perhaps one may still be able to implement the SoV  in some modified way, e.g. making the separated coordinates noncommutative or Grassmannian. We leave this as an important open question for the future.  Curiously, 
we observed\footnote{for the first few values of $L$} that in the standard basis the matrix elements of $\Bg(u)\Bg(v)$ and $\Bg(v)\Bg(u)$ are either equal or are related via multiplication by  $-\frac{u-v-i}{u-v+i}$. This factor is furthermore precisely the one appearing in the commutation relation of the standard $B(u)$ operators in the $\msu(1|1)$ case (see \eq{Bcomms}). The same observation is true for the $\msu(1|1)$ $B^g$ operator we present in appendix \ref{sec:su11}. It would be highly interesting to understand the algebraic implications of these commutation relations.

In the $\msu(n)$ case one could use the same operator $B^g$ to build the states starting from a different reference state, using solutions of the appropriate dual Bethe equations corresponding to a particle-hole transformation in the Bethe ansatz. The $B^g$ operator we constructed here for $\msu(1|2)$ does not have the same property as it annihilates the states which could serve as alternative pseudovacua, namely
\beq
	\ket{0'}=\begin{pmatrix}0\\ \hline 1\\0\end{pmatrix}\otimes\begin{pmatrix}0\\ \hline 1\\0\end{pmatrix}
	\otimes\dots \otimes \begin{pmatrix}0\\ \hline 1\\0\end{pmatrix}\ 
\eeq
and
\beq
	\ket{0''}=\begin{pmatrix}0\\ \hline 0\\1\end{pmatrix}\otimes\begin{pmatrix}0\\ \hline 0\\1\end{pmatrix}
	\otimes\dots \otimes \begin{pmatrix}0\\ \hline 0\\1\end{pmatrix}\ .
\eeq
This property is also related to the fact that the $\Bg$ operator is nilpotent for a general $K$ of the form \eq{K12}, and thus cannot be diagonalized. This serves as another obstacle to implementing the SoV, as for $\msu(n)$ the eigenvectors of $B^g$ play a key role since they define the basis of separated coordinates in which the wavefunction factorizes (see the discussion in section \ref{sec:sov}). However, in the $\msu(1|1)$ case we managed to circumvent this problem by considering a more general $K$ matrix which gives a diagonalizable $B^g$ operator as discussed in appendix \ref{sec:su11}. We hope that this approach may be adapted to the higher rank case, and in addition one could try to use a $K$ matrix with Grassmann elements in the off-diagonal blocks (see section \ref{sec:B11}), though the interpretation of the resulting operator remains to be clarified. In any case, we believe that the very existence of the construction of the states \eq{psi12} for $\msu(1|2)$ is encouraging for the prospect of developing the SoV program in the future.

\subsection{Extension to the $(2|1)$ grading}

While above we discussed the $\msu(1|2)$ spin chains based on the R-matrix with $(1|2)$  grading, one can alternatively consider a spin chain built from the R-matrix with grading chosen as $(2|1)$  (corresponding to $m=2$ and $n=1$ in the notation of section \ref{sec:rev}). This spin chain  
still realizes the  $\msu(1|2)$ symmetry but  differs technically from the case we considered.
Here we present our construction of eigenstates for this choice of grading, which should  also provide important guidance towards its generalization to any $\msu(m|n)$ model.

At the level of transfer matrix eigenvalues, the difference between two choices of the grading can be stated explicitly using the expression for eigenvalues in terms of the Bethe roots, which for the $(2|1)$ case reads\footnote{We have extensively checked this result numerically and it can also be proven using the standard nested Bethe ansatz.}
\beq
\label{T21e}
    \gt(u)=-Q_\theta^-\[\lambda_3\frac{Q_\theta^{---}Q_u}{Q_\theta^- Q_u^{--}}
	-\lambda_2\frac{Q_uQ_v^{---}}{Q_u^{--}Q_v^-}
	-{\lambda_1}\frac{Q_v^{+}}{Q_v^-}\] \ ,
\eeq
where as before we define
\beq
    Q_u=\prod_{j=1}^K(u-u_j) \ , \ \ \ Q_v=\prod_{j=1}^N(u-v_j) \ ,
\eeq
and the Bethe roots $u_i,v_j$ are fixed by standard nested Bethe equations
\beqa
\label{bae21}
   \prod_{k=1}^L\frac{u_i-\theta_k+i/2}{u_i-\theta_k-i/2}&=&\frac{\lambda_3}{\lambda_2}\prod_{j=1}^N\frac{u_i-v_j+i/2}{u_i-v_j-i/2}\ , \ \ \ i=1,\dots,K
   \\ 
   \label{bae21v}
    \prod_{j=1}^K\frac{v_i-u_j+i/2}{v_i-u_j-i/2}&=&-\frac{\lambda_1}{\lambda_2}\prod_{j=1}^N\frac{v_i-v_j+i}{v_i-v_j-i}\ ,\ \ \ i=1,\dots,N \ \ .
\eeqa
Notice that the only difference with the $(1|2)$ Bethe equations \eq{bae12}, \eq{bae122} is a reshuffling of the twists $\lambda_i$. Comparing \eq{T21e} with the eigenvalues of $\gt$ for the $(1|2)$ grading given in \eq{T12e}, we see that the eigenvalues for the two gradings are mapped to each other if we apply complex conjugation supplemented by a shift of $u$, permutation of the twists and overall change of sign,
\beq
    \gt^{(1|2)}(\lambda_1,\lambda_2,\lambda_3,\theta_i,u)=-\[\gt^{(2|1)}(\lambda_3^*,\lambda_2^*,\lambda_1^*,\theta_i^*,u^*+i)\]^* \ .
\eeq
This equality holds at the level of eigenvalues, with a suitable one-to-one identification between eigenvectors in the two models.

Despite the simplicity of this map, there seems to be  no simple relation between $\gt(u)$ (or other entries $T_{ij}(u)$) as operators in the standard basis for the two choices of grading, making the realization of our construction in the $(2|1)$ case nontrivial. However, we found that there still exists a $B^g$ operator which allows one to generate the eigenstates, and it reads
\beq
\label{Bg21}
    \Bg=T_{32}(u+i)\Ber_{31|32}(u+i)+T_{12}(u+i)\Ber_{31|12}(u+i)\ \ ,
     \ \ \text{with}\ \ {T_{ij}\to T^{g}_{ij}}\ ,
\eeq
where now to define $T^{g}=K^{-1}TK$ we should use a matrix $K$ that is generic but has nonzero entries only in the diagonal blocks corresponding to the $(2|1)$ grading,
\beq
\label{K21}
    K=
    \(
    \begin{array}{cc|c}
    K_{11}& K_{12}&0\\
    K_{21}& K_{22}&0\\
    \hline
    0& 0&K_{33}
    \end{array}
    \)\ .
\eeq
This expression should be compared with the $\Bg$ operator for the $(1|2)$ grading given in \eq{Bg12}. We see there is an extra shift of $u$  in the result for the $(2|1)$ case. Up to this shift, if we replace the Berezinians in both results by the usual quantum minors, we find that both expressions are instances of the $\msu(3)$ $\Bg$ operators, simply corresponding to different choices of $K$ for $\msu(3)$. However, for super spin chains only block-diagonal matrices $K$ (like \eq{K12} or \eq{K21}) are allowed, so the $\Bg$ operators cannot be mapped to each other by adjusting $K$. Moreover, the $\Bg$ operator in the $(2|1)$ case has to act on the dual vacuum of the form\footnote{For completeness we note that the $T_{ij}$ operators in the $(2|1)$ case act on the standard reference state $\ket{0}$ defined in \eq{def0} as $T_{11}(u)\ket{0}=\lambda_1Q_\theta^+\ket{0}, \ \ 
	T_{22}(u)\ket{0}=\lambda_2Q_\theta^-\ket{0}, \ \ 
	T_{33}(u)\ket{0}=\lambda_3Q_\theta^-\ket{0}, \ \ 
	T_{23}(u)\ket{0}=0$, and  $T_{ij}(u)\ket{0}=0$ for $i>j$. For the dual vacuum $\ket{0''}$ we have $T_{11}(u)\ket{0''}=\lambda_1Q_\theta^-\ket{0''}, \ \ 
	T_{22}(u)\ket{0''}=\lambda_2Q_\theta^-\ket{0''}, \ \ 
	T_{33}(u)\ket{0''}=\lambda_3Q_\theta^{---}\ket{0''}, \ \ 
	T_{21}(u)\ket{0''}=0$, and  $T_{ij}(u)\ket{0''}=0$ for $i<j$.}
\beq
	\ket{0''}=\begin{pmatrix}0\\0\\ \hline 1\end{pmatrix}\otimes\begin{pmatrix}0\\0\\ \hline 1\end{pmatrix}
	\otimes\dots \otimes \begin{pmatrix}0\\0\\ \hline 1\end{pmatrix}\ .
\eeq
That is rather natural as this is the image of the original $\ket{0}$ vacuum under the map from $\mC^{1|2}$ to $\mC^{2|1}$. The states are built as
\beq
\label{psi21}
    \ket{\Psi}=\Bg(u_1)\dots\Bg(u_K)\ket{0''}
\eeq
where as usual $u_i$ are the momentum-carrying roots fixed by the nested Bethe equations given above in \eq{bae21}, \eq{bae21v}.

As for the $(1|2)$ grading, we have checked numerically that this operator generates the full basis of states  for spin chain length $L=1,2,3$ and $4$, as well as several states  for $L=5$ with up to four magnons. Its other properties also directly parallel the $(1|2)$ case, in particular it is nilpotent, is not suitable for generating states starting from a different vacuum, and the entries of $\Bg(u)\Bg(v)$ are related with those of $\Bg(v)\Bg(u)$ via multiplication by $-\frac{u-v-i}{u-v+i}$ in the cases when they are not equal. Lastly, let us note that the extra shift by $i$ in the result \eq{Bg21} is rather intriguing and is similar to the shift needed for $\msu(1|1)$ when using the $C$ operators to build the states from the dual vacuum as discussed in appendix \ref{sec:su11}.

% While the $\msu(1|2)$ and $\msu(2|1)$ spin chains are of course equivalent, it is instructive to write the explicit result for the $\Bg$ operator in the $\msu(2|1)$ case as well. This also should provide hints towards a generalizaion to any $\msu(m|n)$ model.

% Let us also note that similarly to the $\msu(1|2)$ case, the $\Bg$ operator here is not suitable for building the states starting from a different vacuum, e.g. from the usual $\ket{0}$ reference state.

\section{Conclusions}
\label{sec:concl}

In this paper we presented a new and highly compact construction for the eigenstates of higher-rank supersymmetric rational spin chain, for the first nontrivial example which is $\msu(1|2)$. It is inspired by the analogous proposal in the $\msu(n)$ case which in turn has its roots in the separation of variables approach. We find it rather nontrivial that an analogous construction  of states exists for $\msu(1|2)$  despite the fact that there is no  known implementation of the SoV  in the supersymmetric case. 

While we have checked the proposal extensively, it would be interesting to prove it rigorously, which is likely to be possible in view of the recent proof in the $\msu(3)$ case \cite{Liashyk:2018qfc}. 
It would be also highly  important, though challenging, to uncover the algebraic origins of the operator $\Bg$ which generates the states, and to understand its interpretation within the Yangian. In particular, it would be interesting to understand the algebraic meaning of non-conventional super quantum minors entering our $\msu(1|2)$ construction (see discussion after \eq{Tgr}). A better algebraic understanding would be important for extending our construction to any $\msu(m|n)$, which is one of the key future directions.

While we have focused on the spin chains with a fundamental representation at each site, we hope the construction should work directly for many other representations, as already proven in the $\msu(3)$ case \cite{Liashyk:2018qfc}. It would be important to generalize both the bosonic and supersymmetric constructions to arbitrary representations, in particular to the antisymmetric representation of $\msu(4)$ relevant for 1-point functions in $\cN=4$ SYM with a defect \cite{deLeeuw:2015hxa,deLeeuw:2017cop,deLeeuw:2018mkd}), as well as to noncompact spin chains. Another curious direction is to look for relations with the construction of spin chain Q-operators \cite{Belitsky:2006cp, Frassek:2017bfz,Frassek:2010ga}. It is also interesting to explore deformations of our construction corresponding to the trigonometric XXZ case and to the Gaudin models (either bosonic or supersymmetric \cite{MukhinSup}). In the latter case one may expect an interplay with the remarkable Knizhnik-Zamolodchikov equations \cite{Reshetikhin:1994qw,Ribault:2008si}.

We hope that our results should help to shed light on the yet to be developed SoV program for the supersymmetric case. We present some first steps towards the SoV for  $\msu(1|1)$ spin chains in appendix \ref{sec:su11}. Since in the bosonic case the SoV leads to remarkable results for  correlators (see e.g. \cite{Kitanine:2015jna,Kitanine:2016pvg, Kozlowski:2015ixa}), one may hope for similar simplifications in supersymmetric models. For $\cN=4$ SYM drastic simplification of certain correlators in separated variables was observed very recently in \cite{Cavaglia:2018lxi} (see also \cite{Giombi:2018qox} and \cite{Kazama:2012is,Kazama:2013rya,Kazama:2013qsa,Kazama:2016cfl,Jiang:2015lda,Sobko:2013ema}), making further development of the SoV program all the more important.

\section*{Acknowledgements}
We thank G.~Bednik, J.~Caetano,  A.~Cavaglia, Y.~Jiang, 
Julius,
V.~Kazakov,
I.~Kostov, J.~Lamers,
M.~de~Leeuw,
S.~Leurent,
A.~Liashyk, D.~Serban, G.~Sizov, N.~Slavnov, Z.~Tsuboi and K.~Zarembo for related discussions. N.G. wishes to thank STFC for
support from Consolidated grant number ST/J002798/1. N.G. thanks for hospitality the Laboratoire de Physique Theorique at ENS Paris, where a part of this work was completed.  F.L.-M. was supported by LabEX ENS-ICFP: ANR-10-LABX-0010/ANR-10-IDEX-0001-02 PSL*. F.L.-M. is grateful to Nordita Institute (Stockholm) for hospitality during a part of the work on this project.

\appendix

\section{Comments on $\msu(1|1)$ spin chains}

\label{sec:su11}

In this appendix we discuss the simplest supersymmetric spin chains with $\msu(1|1)$ symmetry. We will see that despite their simplicity it is not clear how to explicitly construct the basis of Sklyanin's separated variables, and we will make some first steps in this direction.

In the $\msu(1|1)$ case, the monodromy matrix $T(u)$ is a $2\times 2$ matrix whose entries $A,B,C,D$ act on the physical  Hilbert space,
\beq
\label{T11}
T(u)=\begin{pmatrix}
	A(u)&B(u) \\ C(u)&D(u)
\end{pmatrix} \ .
\eeq
The transfer matrix is given by its supertrace,
\beq
    \gt(u)=A(u)-D(u) \ ,
\eeq
and defines a commutative family of operators,
\beq
    [\gt(u),\gt(v)]=0 \ .
\eeq
The standard way to construct its eigenstates is by using the $B(u)$ operator as a creation operator on top of the reference state $\ket{0}$ defined by \eq{def0}. The eigenstates are then given by
\beq
\label{psi11}
    \ket{\Psi}=B(u_1)B(u_2)\dots B(u_K)\ket{0}
\eeq
where $u_j$ are the Bethe roots satisfying the $\msu(1|1)$ Bethe equations,
\beq
\label{bae11}
\frac{\lambda_1}{\lambda_2}\prod_{k=1}^L
\frac{u_j-\theta_k+i/2}{u_j-\theta_k-i/2}=1, \ \ j=1,\dots,K
\ \ \ .
\eeq
Let us note that the l.h.s. of \eq{bae11} is the ratio of eigenvalues of $A(u)$ and $D(u)$ on the vector~$\ket{0}$,
\beq
\label{AD0}
    A(u)\ket{0}=\prod_{k=1}^L(u_j-\theta_k+i/2), \ \ \ \ D(u)\ket{0}=\prod_{k=1}^L(u_j-\theta_k-i/2) \ .
\eeq
In order to prove that \eq{psi11} gives an eigenstate of $\gt$ one uses commutation relations between entries of $T(u)$ following from the RTT relation \eq{RTT}. In particular, we have
\beq
	A(u)B(v)=\frac{u-v-i}{u-v}B(v)A(u)+\frac{i}{u-v}B(u)A(v) \ .
\eeq
Moreover   $D(u)$ satisfies exactly the same commutation relation (in contrast to the $\msu(2)$ case). This means that we have a commutation relation between $B$ and the full $\gt=A-D$,
\beq
\label{commTB}
    \gt(u)B(v)=\frac{u-v-i}{u-v}B(v)\gt(u)+\frac{i}{u-v}B(u)\gt(v)
\eeq
Using also that as a consequence of the RTT relation
\beq
\label{Bcomms}
	B(u)B(v)=-\frac{u-v-i}{u-v+i}B(v)B(u) \ ,
\eeq
one can now easily commute $\gt(u)$ through all the $B$ operators in \eq{psi11} until it hits $\ket{0}$ which is its eigenstate due to \eq{AD0}. It is not hard to show that all unwanted terms generated in the process will cancel due to Bethe equations \eq{bae11}, ensuring that $\ket{\Psi}$ is indeed an eigenstate.

Note that the r.h.s. of \eq{bae11} does not include any interaction between the Bethe roots, which are in this sense independent from each other,  and moreover they should all be pairwise distinct. This makes the spin chain somewhat similar to a model of free fermions. It is clear that for a given $L$ the complete set of possible Bethe roots is fixed from \eq{bae11} with $K=L$, and  any particular state is specified by choosing a subset of these roots.

For completeness let us also discuss how to write the analog of the Baxter $T-Q$ equations for $\msu(1|1)$. Introducing the Q-functions
\beq
\label{Q11}
	Q_1=\prod_{j=1}^K(u-u_j), \ \ Q_\theta=\prod_{k=1}^L(u-\theta_k) \ ,
\eeq
we can rewrite the Bethe equations \eq{bae11} as
\beq
	\frac{Q_\theta^+(u_j)}{Q_\theta^-(u_j)}=\frac{\lambda_2}{\lambda_1} \ .
\eeq
Equivalently, we can write the QQ relation
\beq
\label{QQ11}
	\lambda_1Q_\theta^+-\lambda_2Q_\theta^-=(\lambda_1-\lambda_2)Q_1Q_2 \ ,
\eeq
where $Q_2$ is also a polynomial. We see that the l.h.s. of this equation does not depend on the state, and the state is specified simply by selecting $K$ Bethe roots out of the zeros of the l.h.s. These will be the zeros of $Q_1$, while the other zeros of the l.h.s. will be attributed to $Q_2$. One can say that for any particular state $Q_2$ contains those Bethe roots which are not activated for this state.

In terms of the Q-functions the eigenvalue of $\gt(u)$ has the simple form
\beq
\label{Tres}
	\gt=(\lambda_1-\lambda_2)Q_1^{--}Q_2 \ ,
\eeq
and combining this with the QQ relation \eq{QQ11} we get the analog of the
Baxter equation,
\beq
	\gt Q=(\lambda_1Q_\theta^+-\lambda_2Q_\theta^-)Q^{--} \ .
\eeq

\subsection{Separation of variables overview}
\label{sec:sov}

While the construction of eigenstates \eq{psi11} for $\msu(1|1)$ directly parallels the $\msu(2)$ case, a crucial difference is that the $B$ operators no longer commute with each other and instead satisfy a Zamolodchikov-Faddeev type relation \eq{Bcomms}. This prevents immediate realization of Sklyanin's separation of variables program for $\msu(1|1)$ and makes it a nontrivial open question. Let us recall briefly how the SoV works for $\msu(2)$ spin chains. The $B$ operators in that case commute,
\beq
    [B^{\msu(2)}(u),B^{\msu(2)}(v)]=0 \ ,
\eeq
and therefore one can define the commuting operator roots of $B$ denoted as operators $x_k$,
\beq
    B^{\msu(2)}(u)=B_0\prod_{k=1}^L(u-x_k) \ ,
\eeq
where $B_0$ is a constant. The $x_k$ play the role of separated coordinates, and in their common eigenbasis labelled by their eigenvalues $\rx_k$ we have
\beq
    \bra{\rx_1,\dots,\rx_L}B^{\msu(2)}(u)=C\prod_{k=1}^L(u-\rx_k)\bra{\rx_1,\dots,\rx_L}
\eeq
The eigenstates of the transfer matrix can be again built as in $\eq{psi11}$ with the only difference being in the explicit form of Bethe equations satisfied by $u_j$. Then we see that in the common eigenbasis of  $x_k$ the wavefunction factorizes,
\beq
\label{xpsi}
   \bra{\rx_1,\dots,\rx_L}\Psi\rangle =\prod_{k=1}^L(-1)^KQ_1(\rx_k) \ ,
\eeq
where the Q-function $Q_1(u)$ is defined by \eq{Q11} and encodes the Bethe roots. The factorization of the wavefunction into Q-functions in \eq{xpsi} shows that the separation of variables has been achieved in the $\msu(2)$ case.

The main problem for $\msu(1|1)$ models is that the $B$ operators do not commute, so one cannot diagonalize their roots $x_k$ simultaneously, making it unclear how to construct the basis $\bra{\rx_1,\dots,\rx_L}$ of separated coordinates. For $\msu(1|1)$ spin chains there is also another obstacle -- namely, the standard $B$ operator is nilpotent and cannot be diagonalized at all.\footnote{Informally speaking, the reason why $B$ is nilpotent is that  acting with it many times on the reference state $\ket{0}$ we will eventually reach the state where all spins have been flipped, and this state is annihilated by $B$.} In the next section we will show how to resolve at least this problem, serving as a first step towards the construction of the SoV.

\subsection{Improving the $B$ operator}
\label{sec:B11}

As we discussed above, one problem preventing the SoV implementation for $\msu(1|1)$ spin chains is the fact that $B(u)$ is a nilpotent operator and cannot be diagonalized. In fact the same problem is present also in the $\msu(2)$ case where $B$ is  nilpotent as well. There it can be circumvented by redefining the monodromy matrix via an extra similarity transformation with a generic $2\times 2$ constant matrix $K$ acting in the auxiliary space \cite{Gromov:2016itr,Jiang:2015lda},
\beq
\label{Tg}
    T(u)\to T^{g}(u)=K^{-1}T(u)K \ .
\eeq
This transformation removes degeneracy and makes the new $B$ operator diagonalizable, moreover it is a symmetry of the R-matrix and thus preserves all commutation relations (as well as the trace of $T$).

For supersymmetric $\msu(m|n)$ spin chains the transformation \eq{Tg} would only preserve the commutation relations if elements of $K$ in its off-diagonal $m\times n$ and $n\times m$ blocks are treated as Grassmann variables anticommuting also with elements of $T(u)$ in the same blocks. However, the resulting $B$ operator for $\msu(1|1)$ would contain Grassmann variables, making unclear the interpretation of its eigenstates and eigenvectors.
Nevertheless we can use this approach idea as an inspiration and formally consider for $\msu(1|1)$ the new monodromy matrix
\beq
	T^{g}=\begin{pmatrix}
	1&\beta \\ 0&1
	\end{pmatrix}
	\begin{pmatrix}
	A&B \\ C&D
	\end{pmatrix}
	\begin{pmatrix}
	1&\alpha \\ 0&1
	\end{pmatrix}
	=\begin{pmatrix}
	A+\beta C&B+A\alpha+\beta D +\beta C \alpha \\ C&D+C\alpha
	\end{pmatrix}
\eeq
where $\alpha$ and $\beta$ are Grassmann variables commuting with $A,D$ but anticommuting with $B,C$ as well as with each other. Requiring the supertrace of $T$ to be preserved we find $\beta=-\alpha$. Then we can read off the new $B$-operator $B^{g}\equiv T^{g}_{12}$,
so explicitly
\beq
\label{bg11}
	B^{g}=B(u)+\alpha(A(u)-D(u)) \ .
\eeq

The key observation is that one can take $\alpha$ in this equation \eq{bg11} to be a generic complex number rather than a formal Grassmann parameter. The resulting operator will be diagonalizable, and moreover it will still generate the eigenstates! That is, we can again build the eigenstates as
\beq
    \ket{\Psi}=B^{g}(u_1)\dots B^{g}(u_K)\ket{0}
\eeq
The reason for this is that $B^{g}$ satisfies exactly the same commutation relation \eq{commTB} with $\gt$ as $B$ did,
\beq
	\gt(u)B^{g}(v)=\frac{u-v-i}{u-v}B^{g}(v)\gt(u)+\frac{i}{u-v}B^{g}(u)\gt(v) \ ,
\eeq
which one can check explicitly using \eq{commTB}. This immediately means that the $\Bg$ operator is suitable for building eigenstates.

The advantage of using the $\Bg(u)$ operator is that unlike $B(u)$ it can be diagonalized. Its eigenvalues  read\footnote{we checked this explicitly for the first few values of $L$}
\beq
    (\lambda_1-\lambda_2)\alpha
    \prod_{k=1}^L(u-u_k-i s_k), \ \ s_k=\{0,1\} \ .
\eeq
Curiously,  they coincide with eigenvalues of $\gt(u)$ (given in \eq{Tres}) up to a simple and $u$-independent  factor. However, the meaning of this fact is not completely clear yet, especially since the $B^{g}(u)$ operators do not commute for different values of $u$ so their eigenvectors depend on~$u$.

We did not find any simple commutation relation such as \eq{Bcomms} for two  $B^{g}$ operators. However, curiously, we observed that in the standard basis all matrix elements of
$B^{g}(u)B^{g}(v)$ and $B^{g}(v)B^{g}(u)$ are either equal or are related via multiplication by the same factor as in the $B$ commutation relation \eq{Bcomms}.\footnote{We were also able to construct other operators that generate states and satisfy commutation relations of the type \eq{Bcomms} with different nontrivial factors in the r.h.s. (in one case we get operators that simply anticommute).}

Let us also note that the states created by $B^{g}$ do not depend on $\alpha$ at all. The reason for this is that while $\gt$ appears inside the expression for $B^{g}$, on Bethe roots we have
\beq
	\gt(u_i)\ket{0}=0 \ ,
\eeq
which can be easily checked. Thus when we repeatedly act on $\ket{0}$ with $B^{g}$ we can commute   all $\gt$'s to the right of the $B$'s using \eq{commTB} until the $\gt$'s act on the reference state $\ket{0}$ which is annihilated by them, leaving no dependence on $\alpha$.

\subsection{$B^{g}$ and dual roots}

\label{sec:dual}

One can alternatively try to construct the states starting from a different pseudovacuum state where all spins have been flipped,
\beq
\label{def0p}
	\ket{0'}=\begin{pmatrix}0\\1\end{pmatrix}\otimes\begin{pmatrix}0\\1\end{pmatrix}
	\otimes\dots \otimes \begin{pmatrix}0\\1\end{pmatrix}\ .
\eeq
Then in the standard approach one can build the states using the $C$ operator instead of $B$,
\beq
\label{psid1}
    \ket{\Psi}=
	C(v_1+i)\dots C(v_{K'}+i)\ket{0'}\ ,
\eeq
where the dual Bethe roots $v_i$ satisfy the same Bethe equations as before,
\beq
\label{baed}
	\frac{Q_\theta^+(v_j)}{Q_\theta^{-}(v_j)}=\frac{\lambda_2}{\lambda_1}\ , \ \ \ j=1,\dots,K' \ \ \ .
\eeq
We shifted the arguments of $C$ operators by $i$ in \eq{psid1} so as to have the Bethe equations take the conventional form \eq{baed}.
It is not hard to prove that this gives eigenstates, by using the RTT relations as well as
\beq
	A(u)\ket{0'}=Q_\theta^-\ket{0'},\ \ D(u)\ket{0'}=Q_\theta^{---}\ket{0'} \ .
\eeq

In the $\msu(n)$ case one could build the states with $B^{g}$ starting from any of the $n$ dual pseudovacuum states and using the Bethe roots that solve the corresponding dual Bethe equations. However, we found that the operator $\Bg$ we constructed for $\msu(1|1)$ can build states only starting from the usual vacuum $\ket{0}$. The reason for this is that $\Bg$ is a linear combination 
which does not include the $C$ operator, so the dual vacuum $\ket{0'}$ is just an eigenstate of $\Bg(u)$ for all $u$. It would be interesting to see if one may improve the $\Bg$ operator even further, and we leave this question for the future.


\begin{thebibliography}{99}



\bibitem{Sutherland1}
B. Sutherland, ``Beautiful Models: 70 Years of Exactly Solved Quantum Many-Body
Problems''. World Scientific, 2004.


\bibitem{Hubbard1}
F. H. L. Essler, H. Frahm, F. Goehmann, A. Kluemper and V. E. Korepin, ``The
one-dimensional Hubbard model'', Cambridge University Press (2005), Cambridge, UK.

\bibitem{Beisert:2010jr}
  N.~Beisert {\it et al.},
  ``Review of AdS/CFT Integrability: An Overview,''
  Lett.\ Math.\ Phys.\  {\bf 99} (2012) 3
  doi:10.1007/s11005-011-0529-2
  [arXiv:1012.3982 [hep-th]].


\bibitem{Bargheer:2017nne}
  T.~Bargheer, J.~Caetano, T.~Fleury, S.~Komatsu and P.~Vieira,
  ``Handling Handles I: Nonplanar Integrability,''
  arXiv:1711.05326 [hep-th].


\bibitem{Komatsu:2017buu}
  S.~Komatsu,
  ``Lectures on Three-point Functions in N=4 Supersymmetric Yang-Mills Theory,''
  arXiv:1710.03853 [hep-th].


  \bibitem{Basso:2015zoa}
  B.~Basso, S.~Komatsu and P.~Vieira,
  ``Structure Constants and Integrable Bootstrap in Planar N=4 SYM Theory,''
  arXiv:1505.06745 [hep-th].
  

  
  
 
  

\bibitem{Fleury:2016ykk}
  T.~Fleury and S.~Komatsu,
  ``Hexagonalization of Correlation Functions,''
  JHEP {\bf 1701} (2017) 130
  doi:10.1007/JHEP01(2017)130
  [arXiv:1611.05577 [hep-th]].



\bibitem{Cavaglia:2018lxi}
  A.~Cavaglia, N.~Gromov and F.~Levkovich-Maslyuk,
  ``Quantum Spectral Curve and Structure Constants in N=4 SYM: Cusps in the Ladder Limit,''
  arXiv:1802.04237 [hep-th].


\bibitem{Giombi:2018qox}
  S.~Giombi and S.~Komatsu,
  ``Exact Correlators on the Wilson Loop in $\mathcal{N}=4$ SYM: Localization, Defect CFT, and Integrability,''
  arXiv:1802.05201 [hep-th].



\bibitem{SutherlandN1}
B.~Sutherland, ``A General Model for Multicomponent Quantum Systems,'' Phys.
Rev. {\bf B12} (1975) 3795-3805.


\bibitem{Kulish:1983rd}
  P.~P.~Kulish and N.~Y.~Reshetikhin,
  ``Diagonalization Of Gl(n) Invariant Transfer Matrices And Quantum N Wave System (lee Model),''
  J.\ Phys.\ A {\bf 16} (1983) L591.
  doi:10.1088/0305-4470/16/16/001




\bibitem{Belliard:2008di}
  S.~Belliard and E.~Ragoucy,
  ``Nested Bethe ansatz for 'all' closed spin chains,''
  J.\ Phys.\ A {\bf 41} (2008) 295202
  doi:10.1088/1751-8113/41/29/295202
  [arXiv:0804.2822 [math-ph]].



\bibitem{Belliard:2012sn}
  S.~Belliard, S.~Pakuliak, E.~Ragoucy and N.~A.~Slavnov,
  ``Bethe vectors of GL(3)-invariant integrable models,''
  J.\ Stat.\ Mech.\  {\bf 1302} (2013) P02020
  doi:10.1088/1742-5468/2013/02/P02020
  [arXiv:1210.0768 [math-ph]].



\bibitem{Tarasov1994}
Tarasov~V.~O. and Varchenko~A.~N., ``Jackson integral representations for solutions of the
Knizhnik-Zamolodchikov quantum equation'', Algebra i Analiz 6 (1994) 90; St. Petersburg
Math. J. 6 (1995) 275 (Engl. transl.), arXiv:hep-th/9311040.


\bibitem{Currents}
S. Pakuliak, S. Khoroshkin,
{\it The weight function for the quantum af\/f\/ine algebra $U_q(\widehat{\mathfrak{sl}}_3)$},
Theor. Math. Phys. {\bf 145} (2005) 1373, \texttt{math.QA/0610433}.
$\bullet$
S. Khoroshkin, S. Pakuliak, V. Tarasov,
{\it Of\/f-shell Bethe vectors and Drinfeld currents},
J. Geom.  Phys. {\bf 57} (2007) 1713, \texttt{math/0610517}.
$\bullet$
S. Khoroshkin, S. Pakuliak.  {\it A computation of universal weight function for quantum affine algebra $U_q(gl_N)$.} J. Math.  Kyoto Univ. {\bf 48}  (2008) 277, \texttt{math.QA/0711.2819}.
$\bullet$
L. Frappat, S. Khoroshkin, S. Pakuliak, E. Ragoucy, { \it Bethe Ansatz for the Universal Weight Function},
Ann. H. Poincarre {\bf 10} (2009) 513, \texttt{arXiv:0810.3135}.
$\bullet$
A. Oskin, S. Pakuliak, A. Silantyev.
{\it On the universal weight function for the quantum affine
algebra $U_q(\widehat{gl}_{N})$}, St. Petersburg Math. J. {\bf 21} (2010) 651, \texttt{arXiv:0711.2821}.




\bibitem{BelPakR10} S. Belliard, S. Pakuliak, E. Ragoucy, {\it Universal Bethe Ansatz and Scalar Products of Bethe Vectors }, SIGMA {\bf 6} (2010) 94, \texttt{arXiv:1012.1455}.



	\bibitem{Albert:2000ne}
  T.~D.~Albert, H.~Boos, R.~Flume and K.~Ruhlig,
  ``Resolution of the nested hierarchy for rational sl(n) models,''
  J.\ Phys.\ A {\bf 33} (2000) 4963
  doi:10.1088/0305-4470/33/28/302
  [nlin/0002027 [nlin-si]].


\bibitem{Pakuliak:2018rdx}
  S.~Pakuliak, E.~Ragoucy and N.~Slavnov,
  ``Nested Algebraic Bethe Ansatz in integrable models: recent results,''
  arXiv:1803.00103 [math-ph].




\bibitem{Aganagic:2017gsx}
  M.~Aganagic and A.~Okounkov,
  ``Quasimap counts and Bethe eigenfunctions,''
  arXiv:1704.08746 [math-ph].
  

\bibitem{Gromov:2016itr}
  N.~Gromov, F.~Levkovich-Maslyuk and G.~Sizov,
  ``New Construction of Eigenstates and Separation of Variables for SU(N) Quantum Spin Chains,''
  JHEP {\bf 1709} (2017) 111
  doi:10.1007/JHEP09(2017)111
  [arXiv:1610.08032 [hep-th]].



 \bibitem{Liashyk:2018qfc}
  A.~Liashyk and N.~A.~Slavnov,
  ``On Bethe vectors in $\mathfrak{gl}_3$-invariant integrable models,''
  JHEP {\bf 1806} (2018) 018
  doi:10.1007/JHEP06(2018)018
  [arXiv:1803.07628 [math-ph]].


\bibitem{Belliard:2018pie}
  S.~Belliard and N.~A.~Slavnov,
  ``A note on $\mathfrak{gl}_2$-invariant Bethe vectors,''
  arXiv:1802.07576 [math-ph].



\bibitem{Belliard:2018pvg}
  S.~Belliard, N.~A.~Slavnov and B.~Vallet,
  ``Modified Algebraic Bethe Ansatz: Twisted XXX Case,''
  SIGMA {\bf 14} (2018) 054
  doi:10.3842/SIGMA.2018.054
  [arXiv:1804.00597 [math-ph]].


\bibitem{FioravNew}
  D.~Fioravanti and M.~Rossi,
  ``From the braided to the usual Yang-Baxter relation,''
  J.\ Phys.\ A {\bf 34} (2001) L567
  doi:10.1088/0305-4470/34/42/102
  [hep-th/0107050].


\bibitem{Sklyaninold}
E.~K.~Sklyanin, “New approach to the quantum nonlinear Schrod
inger equation”, J.~Phys.~A
22(17), 3551

\bibitem{Gohmann:1999av}
  F.~Gohmann and V.~E.~Korepin,
  ``Solution of the quantum inverse problem,''
  J.\ Phys.\ A {\bf 33} (2000) 1199
  doi:10.1088/0305-4470/33/6/308
  [hep-th/9910253].


\bibitem{Essler:1992he}
  F.~H.~L.~Essler and V.~E.~Korepin,
  ``Higher conservation laws and algebraic Bethe ansatze for the supersymmetric t-J model,''
  Phys.\ Rev.\ B {\bf 46} (1992) 9147.
  doi:10.1103/PhysRevB.46.9147


\bibitem{Foerster:1992uk}
  A.~Foerster and M.~Karowski,
  ``Algebraic properties of the Bethe ansatz for an spl(2,1) supersymmetric t-J model,''
  Nucl.\ Phys.\ B {\bf 396} (1993) 611.
  doi:10.1016/0550-3213(93)90665-C


\bibitem{Gohmann:2001wh}
  F.~Gohmann,
  ``Algebraic Bethe ansatz for the gl(1|2) generalized model and Lieb-Wu equations,''
  Nucl.\ Phys.\ B {\bf 620} (2002) 501
  doi:10.1016/S0550-3213(01)00497-7
  [cond-mat/0108486].



\bibitem{ZhangCM}  F.C.~Zhang, T.M.~Rice, ``Effective Hamiltonian for the superconducting Cu oxides'', Phys.\ Rev.\ B {\bf 37} (1988) 3759–3761.


\bibitem{SchlottmannCM}
P.~Schlottmann, ``Integrable narrow-band model with possible relevance to heavy Fermion
systems'', Phys.\ Rev.\ B {\bf 36} (1987) 5177–5185.


\bibitem{Belliard:2012av}
  S.~Belliard, S.~Pakuliak, E.~Ragoucy and N.~A.~Slavnov,
  ``Form factors in SU(3)-invariant integrable models,''
  J.\ Stat.\ Mech.\  {\bf 1304} (2013) P04033
  doi:10.1088/1742-5468/2013/04/P04033
  [arXiv:1211.3968 [math-ph]].





\bibitem{Belliard:2012pr}
  S.~Belliard, S.~Pakuliak, E.~Ragoucy, N.~A.~Slavnov, S.~Pakuliak, E.~Ragoucy and N.~A.~Slavnov,
  ``Algebraic Bethe ansatz for scalar products in SU(3)-invariant integrable models,''
  J.\ Stat.\ Mech.\  {\bf 1210} (2012) P10017
  doi:10.1088/1742-5468/2012/10/P10017
  [arXiv:1207.0956 [math-ph]].


\bibitem{Belliard:2012is}
  S.~Belliard, S.~Pakuliak, E.~Ragoucy and N.~A.~Slavnov,
  ``Highest coefficient of scalar products in SU(3)-invariant integrable models,''
  J.\ Stat.\ Mech.\  {\bf 1209} (2012) P09003
  doi:10.1088/1742-5468/2012/09/P09003
  [arXiv:1206.4931 [math-ph]].



\bibitem{Pakuliak:2014ela}
  S.~Z.~Pakuliak, E.~Ragoucy and N.~A.~Slavnov,
  ``Determinant representations for form factors in quantum integrable models with the GL(3)-invariant R-matrix,''
  Theor.\ Math.\ Phys.\  {\bf 181} (2014) no.3,  1566
  doi:10.1007/s11232-014-0236-0
  [arXiv:1406.5125 [math-ph]].



\bibitem{Pakuliak:2015qga}
  S.~Pakuliak, E.~Ragoucy and N.~A.~Slavnov,
  ``GL(3)-Based Quantum Integrable Composite Models. I. Bethe Vectors,''
  SIGMA {\bf 11} (2015) 063
  doi:10.3842/SIGMA.2015.063
  [arXiv:1501.07566 [math-ph]].


\bibitem{Pakuliak:2015fma}
  S.~Pakuliak, E.~Ragoucy and N.~A.~Slavnov,
  ``GL(3)-Based Quantum Integrable Composite Models. II. Form Factors of Local Operators,''
  SIGMA {\bf 11} (2015) 064
  doi:10.3842/SIGMA.2015.064
  [arXiv:1502.01966 [math-ph]].



\bibitem{Slavnov:2015qoa}
  N.~A.~Slavnov,
  ``Scalar products in GL(3)-based models with trigonometric R-matrix. Determinant representation,''
  J.\ Stat.\ Mech.\  {\bf 1503} (2015) no.3,  P03019
  doi:10.1088/1742-5468/2015/03/P03019
  [arXiv:1501.06253 [math-ph]].


\bibitem{SlavnovDet}
N.~A.~Slavnov, ``Calculation of scalar products of wave functions and form factors in the
framework of the algebraic Bethe ansatz'', Theor. Math. Phys. {\bf 79}:2 (1989) 502–508.



\bibitem{Slavnov:2016nhu}
  N.~A.~Slavnov,
  ``Multiple commutation relations in the models with $\mathfrak gl(2|1)$ symmetry,''
  Theor.\ Math.\ Phys.\  {\bf 189} (2016) no.2,  1624
   [Teor.\ Mat.\ Fiz.\  {\bf 189} (2016) no.2,  256]
  doi:10.1134/S0040577916110076
  [arXiv:1604.05343 [math-ph]].



\bibitem{Hutsalyuk:2016ndz}
  A.~Hutsalyuk, A.~Liashyk, S.~Z.~Pakuliak, E.~Ragoucy and N.~A.~Slavnov,
  ``Scalar products of Bethe vectors in models with $\mathfrak{gl}(2|1)$ symmetry 1. Super-analog of Reshetikhin formula,''
  J.\ Phys.\ A {\bf 49} (2016) no.45,  454005
  doi:10.1088/1751-8113/49/45/454005
  [arXiv:1605.09189 [math-ph]].


\bibitem{Hutsalyuk:2016yii}
  A.~Hutsalyuk, A.~Liashyk, S.~Z.~Pakuliak, E.~Ragoucy and N.~A.~Slavnov,
  ``Scalar products of Bethe vectors in models with $\mathfrak{g}\mathfrak{l}(2|1)$ symmetry 2. Determinant representation,''
  J.\ Phys.\ A {\bf 50} (2017) no.3,  034004
  doi:10.1088/1751-8121/50/3/034004
  [arXiv:1606.03573 [math-ph]].


\bibitem{Hutsalyuk:2016jwh}
  A.~Hutsalyuk, A.~Liashyk, S.~Z.~Pakuliak, E.~Ragoucy and N.~A.~Slavnov,
  ``Form factors of the monodromy matrix entries in gl(2|1)-invariant integrable models,''
  Nucl.\ Phys.\ B {\bf 911} (2016) 902
  doi:10.1016/j.nuclphysb.2016.08.025
  [arXiv:1607.04978 [math-ph]].


\bibitem{Hutsalyuk:2016gkn}
  A.~Hutsalyuk, A.~Liashyk, S.~Z.~Pakuliak, E.~Ragoucy and N.~A.~Slavnov,
  ``Multiple Actions of the Monodromy Matrix in $\mathfrak{gl}(2|1)$-Invariant Integrable Models,''
  SIGMA {\bf 12} (2016) 099
  doi:10.3842/SIGMA.2016.099
  [arXiv:1605.06419 [math-ph]].


\bibitem{Hutsalyuk:2017tcx}
  A.~Hutsalyuk, A.~Liashyk, S.~Z.~Pakuliak, E.~Ragoucy and N.~A.~Slavnov,
  ``Scalar products of Bethe vectors in the models with $\mathfrak{gl}(m|n)$ symmetry,''
  Nucl.\ Phys.\ B {\bf 923} (2017) 277.
  doi:10.1016/j.nuclphysb.2017.07.020
  

\bibitem{Pakuliak:2016bhc}
  S.~Z.~Pakuliak, E.~Ragoucy and N.~A.~Slavnov,
  ``Bethe vectors for models based on the super-Yangian $Y(\mathfrak{gl}(m|n))$,''
  J. Integrab. Syst. 2 (2017) 1--31
  doi:10.1093/integr/xyx001
  [arXiv:1604.02311 [math-ph]].



\bibitem{Hutsalyuk:2016srn}
  A.~Hutsalyuk, A.~Liashyk, S.~Z.~Pakuliak, E.~Ragoucy and N.~A.~Slavnov,
  ``Bethe vectors in integrable models based on the super-Yangian $Y(\mathfrak{gl}(m|n))$,''
  doi:10.1070/RM9754
  arXiv:1611.09620 [math-ph].


\bibitem{Fuksa:2017jbl}
  J.~Fuksa and N.~A.~Slavnov,
  ``Form factors of local operators in supersymmetric quantum integrable models,''
  J.\ Stat.\ Mech.\  {\bf 1704} (2017) no.4,  043106
  doi:10.1088/1742-5468/aa6686
  [arXiv:1701.05866 [math-ph]].




\bibitem{Hutsalyuk:2017way}
  A.~Hutsalyuk, A.~Liashyk, S.~Z.~Pakuliak, E.~Ragoucy and N.~A.~Slavnov,
  ``Norm of Bethe vectors in models with gl(m|n) symmetry,''
  Nucl.\ Phys.\ B {\bf 926} (2018) 256
  doi:10.1016/j.nuclphysb.2017.11.006
  [arXiv:1705.09219 [math-ph]].



\bibitem{Sklyanin:1992sm}
  E.~K.~Sklyanin,
  ``Separation of variables in the quantum integrable models related to the Yangian Y[sl(3)],''
  J.\ Math.\ Sci.\  {\bf 80} (1996) 1861
   [Zap.\ Nauchn.\ Semin.\  {\bf 205} (1993) 166]
  doi:10.1007/BF02362784
  [hep-th/9212076].

\bibitem{Sklyanin:1995bm}
  E.~K.~Sklyanin,
  ``Separation of variables - new trends,''
  Prog.\ Theor.\ Phys.\ Suppl.\  {\bf 118} (1995) 35
  doi:10.1143/PTPS.118.35
  [solv-int/9504001].

\bibitem{Ltoapp}
A.~Liashyk, in preparation


\bibitem{Frappat:1996pb}
  L.~Frappat, P.~Sorba and A.~Sciarrino,
  ``Dictionary on Lie superalgebras,''
  hep-th/9607161.
  

\bibitem{Volin:2010cq}
  D.~Volin,
  ``Quantum integrability and functional equations: Applications to the spectral problem of AdS/CFT and two-dimensional sigma models,''
  J.\ Phys.\ A {\bf 44} (2011) 124003
  doi:10.1088/1751-8113/44/12/124003
  [arXiv:1003.4725 [hep-th]].


  \bibitem{Zhang:1994ad}
  R.~b.~Zhang,
  ``Representations of superYangian,''
  J.\ Math.\ Phys.\  {\bf 36} (1995) 3854
  doi:10.1063/1.530932
  [hep-th/9411243].


\bibitem{Zhang:1995uh}
  R.~b.~Zhang,
  ``The gl(m|n) superYangian and its finite dimensional representations,''
  Lett.\ Math.\ Phys.\  {\bf 37} (1996) 419
  doi:10.1007/BF00312673
  [q-alg/9507029].
  
  

\bibitem{MolevBook}
A.~Molev.  Yangians and Classical Lie Algebras.
Mathematical Surveys and Monographs 143, AMS, Providence, RI, 2007.
  
  

\bibitem{Berezin87}
Felix~ Alexandrovich~ Berezin. Introduction to superanalysis, volume 9 of Mathematical
Physics and Applied Mathematics. D. Reidel Publishing Co., Dordrecht,
1987. Edited and with a foreword by A. A. Kirillov, With an appendix by
V. I. Ogievetsky, Translated from the Russian by J. Niederle and R. Kotecky,
Translation edited by Dimitri Leites.



\bibitem{FioresiBook}
Carmeli, C., Caston, L., and Fioresi, R. (2011). Mathematical foundations of supersymmetry. European Mathematical Society.



\bibitem{NazarovBer}
Nazarov, M. L. ``Quantum Berezinian and the classical Capelli identity.'' Letters in Mathematical Physics 21.2 (1991): 123-131.


\bibitem{Stukopin94}
V. A. Stukopin. Yangians of Lie superalgebras of type A(m, n). Functional
Analysis and its Applications, 28(3):217–219, 1994.


\bibitem{Gow05}
Gow, Lucy. "On the Yangian $Y (\mathfrak{g}\mathfrak{l} _ {m| n})$ and its quantum Berezinian." Czechoslovak Journal of Physics 55.11 (2005): 1415-1420.


\bibitem{Gow07}
Gow, Lucy. "Gauss decomposition of the Yangian." Communications in Mathematical Physics 276.3 (2007): 799-825.


\bibitem{GowThesis}
Gow, Lucy. Yangians of Lie superalgebras. Diss. The University of Sydney, 2007.


  \bibitem{deLeeuw:2015hxa}
  M.~de Leeuw, C.~Kristjansen and K.~Zarembo,
  ``One-point Functions in Defect CFT and Integrability,''
  JHEP {\bf 1508} (2015) 098
  doi:10.1007/JHEP08(2015)098
  [arXiv:1506.06958 [hep-th]].
  

  \bibitem{deLeeuw:2017cop}
  M.~de Leeuw, A.~C.~Ipsen, C.~Kristjansen and M.~Wilhelm,
  ``Introduction to Integrability and One-point Functions in $\mathcal{N}=4$ SYM and its Defect Cousin,''
  arXiv:1708.02525 [hep-th].
  
  

  \bibitem{deLeeuw:2018mkd}
  M.~De Leeuw, C.~Kristjansen and G.~Linardopoulos,
  ``Scalar one-point functions and matrix product states of AdS/dCFT,''
  Phys.\ Lett.\ B {\bf 781} (2018) 238
  doi:10.1016/j.physletb.2018.03.083
  [arXiv:1802.01598 [hep-th]].
  
\bibitem{Frassek:2017bfz}
  R.~Frassek, C.~Marboe and D.~Meidinger,
  ``Evaluation of the operatorial Q-system for non-compact super spin chains,''
  JHEP {\bf 1709} (2017) 018
  doi:10.1007/JHEP09(2017)018
  [arXiv:1706.02320 [hep-th]].


\bibitem{Belitsky:2006cp}
  A.~V.~Belitsky, S.~E.~Derkachov, G.~P.~Korchemsky and A.~N.~Manashov,
  ``Baxter Q-operator for graded SL(2|1) spin chain,''
  J.\ Stat.\ Mech.\  {\bf 0701} (2007) P01005
  doi:10.1088/1742-5468/2007/01/P01005
  [hep-th/0610332].


\bibitem{Frassek:2010ga}
  R.~Frassek, T.~Lukowski, C.~Meneghelli and M.~Staudacher,
  ``Oscillator Construction of su(n|m) Q-Operators,''
  Nucl.\ Phys.\ B {\bf 850} (2011) 175
  doi:10.1016/j.nuclphysb.2011.04.008
  [arXiv:1012.6021 [math-ph]].
  

  \bibitem{MukhinSup}
 E.~Mukhin,  B.~Vicedo,  C.A.S.~Young, ``Gaudin models for $gl(m|n)$'', J.Math.Phys. 56 (5), 05170


  \bibitem{Reshetikhin:1994qw}
  N.~Reshetikhin and A.~Varchenko,
  ``Quasiclassical asymptotics of solutions to the KZ equations,''
  hep-th/9402126.
  

  \bibitem{Ribault:2008si}
  S.~Ribault,
  ``On sl(3) Knizhnik-Zamolodchikov equations and W(3) null-vector equations,''
  JHEP {\bf 0910} (2009) 002
  doi:10.1088/1126-6708/2009/10/002
  [arXiv:0811.4587 [hep-th]].
  
  

\bibitem{Kitanine:2015jna}
  N.~Kitanine, J.~M.~Maillet, G.~Niccoli and V.~Terras,
  ``On determinant representations of scalar products and form factors in the SoV approach: the XXX case,''
  J.\ Phys.\ A {\bf 49} (2016) no.10,  104002
  doi:10.1088/1751-8113/49/10/104002
  [arXiv:1506.02630 [math-ph]].

\bibitem{Kitanine:2016pvg}
  N.~Kitanine, J.~M.~Maillet, G.~Niccoli and V.~Terras,
  ``The open XXX spin chain in the SoV framework: scalar product of separate states,''
  J.\ Phys.\ A {\bf 50} (2017) no.22,  224001
  doi:10.1088/1751-8121/aa6cc9
  [arXiv:1606.06917 [math-ph]].
  
 \bibitem{Kozlowski:2015ixa}
  K.~K.~Kozlowski,
  ``Asymptotic analysis and quantum integrable models,''
  arXiv:1508.06085 [math-ph]. 


  \bibitem{Kazama:2012is}
  Y.~Kazama and S.~Komatsu,
  ``Wave functions and correlation functions for GKP strings from integrability,''
  JHEP {\bf 1209} (2012) 022
  doi:10.1007/JHEP09(2012)022
  [arXiv:1205.6060 [hep-th]].
  

  \bibitem{Kazama:2013rya}
  Y.~Kazama, S.~Komatsu and T.~Nishimura,
  ``A new integral representation for the scalar products of Bethe states for the XXX spin chain,''
  JHEP {\bf 1309} (2013) 013
  doi:10.1007/JHEP09(2013)013
  [arXiv:1304.5011 [hep-th]].
  

  \bibitem{Kazama:2013qsa}
  Y.~Kazama and S.~Komatsu,
  ``Three-point functions in the SU(2) sector at strong coupling,''
  JHEP {\bf 1403} (2014) 052
  doi:10.1007/JHEP03(2014)052
  [arXiv:1312.3727 [hep-th]].
  
  

\bibitem{Kazama:2016cfl}
  Y.~Kazama, S.~Komatsu and T.~Nishimura,
  ``Classical Integrability for Three-point Functions: Cognate Structure at Weak and Strong Couplings,''
  JHEP {\bf 1610} (2016) 042
   Erratum: [JHEP {\bf 1802} (2018) 047]
  doi:10.1007/JHEP10(2016)042, 10.1007/JHEP02(2018)047
  [arXiv:1603.03164 [hep-th]].
  

\bibitem{Jiang:2015lda}
  Y.~Jiang, S.~Komatsu, I.~Kostov and D.~Serban,
  ``The hexagon in the mirror: the three-point function in the SoV representation,''
  J.\ Phys.\ A {\bf 49} (2016) no.17,  174007
  doi:10.1088/1751-8113/49/17/174007
  [arXiv:1506.09088 [hep-th]].




  \bibitem{Sobko:2013ema}
  E.~Sobko,
  ``A new representation for two- and three-point correlators of operators from sl(2) sector,''
  JHEP {\bf 1412} (2014) 101
  doi:10.1007/JHEP12(2014)101
  [arXiv:1311.6957 [hep-th]].
  
\end{thebibliography}
\end{document}